\documentclass[twocolumn,showpacs,preprintnumbers,amsmath,amssymb,superscriptaddress,prl]{revtex4-2}

\usepackage{comment}
\usepackage{graphicx}
\usepackage[]{hyperref}      
\hypersetup{
    colorlinks,
    linkcolor={black},
    citecolor={blue},
    urlcolor={blue!80!black}
}

\usepackage{xcolor}

\usepackage{ifthen}
\usepackage{expl3}

\newcommand{\processdigit}[1]{%
  \ifthenelse{\equal{#1}{1}}{$\blacksquare$\hspace{0.0pt}}{$\square$\hspace{0.0pt}}%
}

\makeatletter
\newcommand{\config}[1]{%
\mbox{\@tfor\next:=#1\do{\processdigit{\next}}%
  \hspace{-3.0pt}}
}
\makeatother

\allowdisplaybreaks

\begin{document}

\title{Universal fragmentation in annihilation reactions with constrained kinetics}

\author{Enrique Rozas Garcia}
\email{enrique.rozas.garcia@physics.gu.se}
\affiliation{Department of Physics, Gothenburg University, 41296 Gothenburg, Sweden}

\author{Alfred Weddig Karlsson}
\email{weddig@student.chalmers.se}
\affiliation{Department of Physics, Chalmers University of Technology, 41296 Gothenburg, Sweden}

\author{Johannes Hofmann}
\email{johannes.hofmann@physics.gu.se}
\affiliation{Department of Physics, Gothenburg University, 41296 Gothenburg, Sweden}
\affiliation{Nordita, Stockholm University and KTH Royal Institute of Technology, 10691 Stockholm, Sweden}

\date{\today}

\begin{abstract}
In reaction-diffusion models of annihilation reactions in low dimensions, single-particle dynamics provides a bottleneck for reactions, leading to an anomalously slow approach to the empty state. Here, we construct a reaction model with a reciprocal bottleneck of reactions on particle dynamics in which single-particle motion conserves the center of mass. We show that such a constrained reaction-diffusion dynamics does not approach an empty state but freezes at late times in a state with fragmented particle clusters. The late-time dynamics and final density are universal, and we provide exact results for the final density in the large-reaction rate limit. Thus, our setup constitutes a minimal model for the fragmentation of a one-dimensional lattice into independent particle clusters. We suggest that the universal reaction dynamics could be observable in experiments with cold atoms or in the Auger recombination of exciton gases.
\end{abstract}

\maketitle  

Reaction-diffusion models describe a plethora of biological, chemical, and physical systems~\cite{vankampen07,krapivsky13}, ranging from population dynamics~\cite{volpert2009reaction}, colloid clustering~\cite{meakin1988models,sciortino2005glassy},  reactions in polymer melts~\cite{degennes82}, models of the early universe~\cite{toussaint83}, to spin dynamics~\cite{family91} and exciton recombination in semiconductors~\cite{russo06,allam13}. In their simplest form, they describe the stochastic annihilation \mbox{$A + A \stackrel{\lambda}{\rightarrow} \emptyset$} of two reactants $A$ with a rate $\lambda$. At late times, the average reactant density $\rho_t$ approaches the empty state with a mean-field decay \mbox{$\rho_t \rightarrow 1/(\lambda t)$}, which is universal in the sense that it does not depend on the initial reactant number. This universality is pushed even further if the reactants are allowed to diffuse in dimensions \mbox{$d \leq 2$}~\cite{krapivsky13}. Here, local reactions create depletion zones, which must be closed by diffusion before further annihilations can take place. The late-time decay of the density is then limited by the diffusion timescale across a depletion zone, resulting in a slower power-law decay \mbox{$\rho_t \rightarrow 1/\sqrt{D t}$} that depends on the diffusion constant~$D$ but not on the reaction rate~$\lambda$~\cite{cardy96}. Modifications of the particle dynamics---for example, for sub- or superdiffusive dynamics~\cite{chen01,yuste02,sokolov06} or quantum random walks~\cite{perfetto23}---result in different power-law decays to the final empty state. Fundamentally, the single-particle dynamics acts as a bottleneck for particle reactions.

In this Letter, we study  how the late-time dynamics of reaction-diffusion systems changes when the single-particle dynamics is constrained by conservation laws. To this end, we introduce a model where the motion of a particle is tied to that of its neighbors through center-of-mass conservation, which creates a reciprocal bottleneck for the low-density particle dynamics. We find that the interplay between reactions and constrained dynamics fragments the system into independent regions as the density decreases, which leads to a nonzero final density as opposed to an empty state. In particular, both the late-time dynamics and the asymptotic density are universal. This is distinct from simple diffusive dynamics where transport limits reactions but reactions do not impact transport. Constrained kinetics, without the inclusion of reactions, have been recently discussed as a model for fracton dynamics~\cite{han23, Feldmeier20}, and may lead to the breakdown of thermalization in quantum systems due to the fragmentation of Hilbert space \cite{rakovszky2020statistical, sala2020ergodicity, khemani2020localization, Morningstar20, pozderac2023exact}. Experimentally, they have been realized with ultracold quantum gases in disorder potentials~\cite{martirosyan23} and in optical lattices~\cite{guardadosanchez20}. Here, the combination with annihilation reactions provides a simple model where the correlated dynamics freezes out at late times.

\begin{figure}[b]
\includegraphics[width=0.9\linewidth]{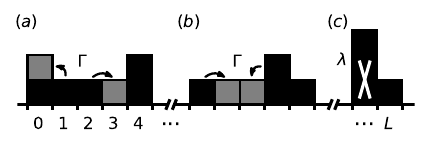}
\caption{Sketch of events in our model.
Black boxes represent particles, and gray boxes locations after an update. 
Pairs of particles can either (a) hop outward from adjacent sites with rate~$\Gamma$, (b) execute the reverse inward hop with equal rate, or (c) annihilate at the same lattice site with rate~$\lambda$.}
\label{fig:1}
\end{figure}

In detail, we consider correlated hopping steps on a one-dimensional lattice, where pairs of adjacent particles hop one lattice site apart with rate $\Gamma$. This is illustrated in Fig.~\ref{fig:1}(a), where particles at the adjacent sites~$1$ and~$2$ simultaneously hop outward to sites $0$ and $3$, marked in gray. We also allow the reverse inward hopping step with equal rate [Fig.~\ref{fig:1}(b)]. Annihilation reactions remove a pair of particles on the same lattice site with rate $\lambda$ [Fig.~\ref{fig:1}(c)]. We simulate the time evolution of this model on a periodic lattice of length~$L$ using the Gillespie algorithm~\cite{gillespie2007stochastic}, where the initial particle number at each site is Poisson distributed with mean $\rho_0$ and we use \mbox{$L=10^6$} unless stated otherwise. For each iteration, the Gillespie algorithm determines the exponentially distributed time to the next event from the total rate of possible correlated hopping and annihilation steps, decides on a random annihilation or hopping step, and updates the state accordingly. To efficiently keep track of the possible correlated hopping and reaction steps between iterations, we  modify a bookkeeping algorithm---introduced in Ref.~\cite{park05} to simulate bosonic reaction-diffusion systems---to include the constrained dynamics~\cite{weddigkarlsson23}. We stop the time evolution when a final state is reached, i.e., a state for which the density remains constant at all later times. 

\begin{figure}[t!]
\includegraphics{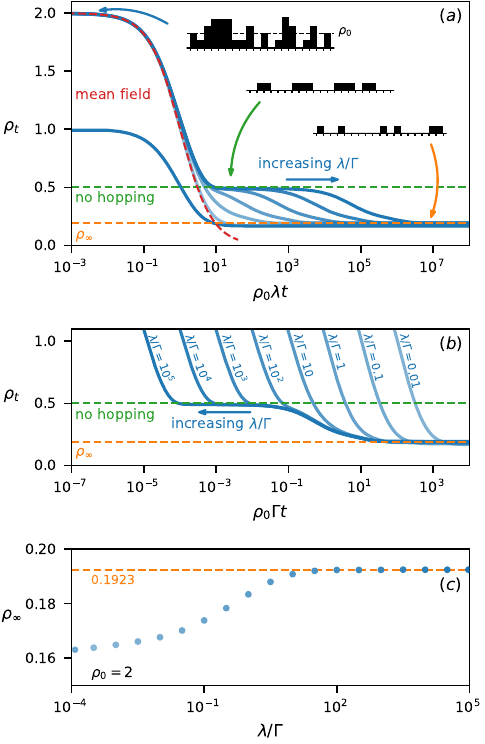}
\caption{ (a) Time evolution of the particle density~$\rho_t$ for various initial densities~$\rho_0$ and ratios \mbox{$\lambda/\Gamma$}. The density does not decrease to zero at late times but arrests at a finite value. Top inset: Particles in the initial state are Poisson distributed with average particle number~$\rho_0$. Middle inset: In the absence of particle dynamics, the final state is the parity projection of the initial distribution. This state is reached as a transient for \mbox{$\lambda\gg\Gamma$} (green dashed line). Bottom inset: Final state for the constrained dynamics. In this state, correlated hopping steps are possible but do not lead to reactions. (b) Same data as (a) as a function of $\rho_0 \Gamma t$, which showcases late-time universality. (c) Final density $\rho_\infty$ for \mbox{$\rho_0=2$} as a function of $\lambda/\Gamma$, averaged over $50$ runs. The variance is too small to be visible.
}
\label{fig:2}
\end{figure}

Figure~\ref{fig:2}(a) shows the time evolution of the average density $\rho_t$ as a function of the scaling variable $\rho_0 \lambda t$ for different initial densities \mbox{$\rho_0$} and ratios of reaction to hopping rates \mbox{$\lambda/\Gamma$}. The top inset illustrates a typical Poisson-distributed initial configuration. The early-time evolution shows a scaling collapse to the  $\Gamma$-independent mean-field result \mbox{$\rho_t = \rho_0/(1+\rho_0 \lambda t)$} (dashed red line), indicating a reaction-dominated regime where correlated particle motion does not significantly affect the density. In the absence of correlated hopping, \mbox{$\Gamma=0$}, the final state is the parity projection of the initial Poisson-distributed state (middle inset), described by a Bernoulli distribution with mean \mbox{$\rho = e^{-\rho_0} \sum_{n=0}^{\infty} \rho_0^{2n}/(2n)! = \frac{1}{2}\left(1-e^{-2\rho_0}\right)$}, which is approximately equal to~$1/2$ for sufficiently large~$\rho_0$ (green dashed line). For large reaction rates, \mbox{$\lambda\gg\Gamma$}, this state is reached as a transient to the asymptotic final state, after which the correlated particle dynamics mixes the state and enables a further reduction in the density. However, unlike for simple diffusion dynamics, as the particle density decreases the state fragments into separate regions, which hinders the transport of multiple particles to equal lattice sites. Eventually, no further annihilations are possible and the density arrests at a nonzero value $\rho_\infty$ (dashed orange line, bottom inset). A main result of our work is that, for high enough initial densities, the final density is approximately independent of $\rho_0$, $\lambda$, and~$\Gamma$. 

Whether a configuration is maximally fragmented and cannot decay to a state with a smaller particle number is determined by the presence of a small number of short particle sequences. Indeed, in order to check if a state is final, we first verify that there is at most one particle per lattice site, i.e., that no annihilation can take place in the next step. In this case, we check for groups of particles that can react after one correlated hopping step; these must include at least one of the sequences \mbox{\config{111},} \mbox{\config{1101},} or \mbox{\config{1011},} where a black (white) box indicates an occupied (empty) site. If these sequences are also absent, we successively check for the smallest sequences that can lead to annihilation after two \mbox{(\config{10011}}, \mbox{\config{11001})}, three \mbox{(\config{101001},} \mbox{\config{100101},} \mbox{\config{1100011})}, and four \mbox{(\config{10100011},} \mbox{\config{11000101})} hopping steps. Here, the longer sequences are constructed from shorter ones by correlated hopping steps. This process terminates at five hops, since all possible five-hop sequences contain a smaller subsequence that can react in fewer steps. Thus, a state is final if and only if it does not contain any of the ten particle sequences depicted above.

\begin{figure}[t]
    \includegraphics{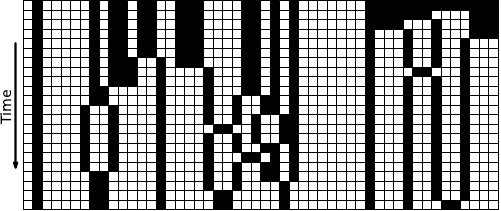}
    \caption{
    Illustration of the instantaneous-reaction dynamics for \mbox{$L=50$}. Black (white) boxes indicate occupied (empty) lattice sites and time evolves from the top to the bottom. Shown are the configurations after each update.}
\label{fig:3}
\end{figure}

To demonstrate that the late-time dynamics is universal and determined by the correlated hopping, we show in Fig.~\ref{fig:2}(b) the scaling collapse of the density onto a single curve that depends on the  dimensionless time \mbox{$\rho_0 \Gamma t$}. This is shown in more detail in Fig.~\ref{fig:2}(c), which shows the final density $\rho_\infty$ as a function of the ratio \mbox{$\lambda/\Gamma$} for \mbox{$\rho_0=2$}. In the limit of fast reactions, \mbox{$\lambda\gg \Gamma$}, the final density arrests at a universal value \mbox{$\rho_\infty = 0.1923(2)$}, which is indicated by the dashed orange line. In this limit, there is a clear separation between the initial reaction-limited dynamics, which first relaxes to the transient state, and the transport-limited dynamics that evolve to the final state. At finite reaction rates, there are small corrections to this result, as the presence of multiply occupied sites aids mixing, resulting in a smaller final density [Fig.~\ref{fig:2}(c)].

\begin{figure}[t]
    \includegraphics{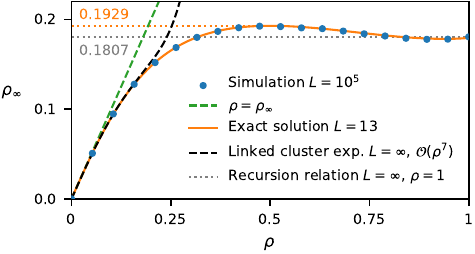}
    \caption{
    Average final density~$\rho_\infty$ with instantaneous reactions for Bernoulli-distributed initial states with density~$\rho$. Numerical results (blue points) are averaged over $50$ samples with \mbox{$L=10^5$}; the variance is too small to be displayed. The exact final density for a system of size \mbox{$L=13$} [orange line, Eq.~\eqref{eq:poly}] agrees well with the numerics, illustrating fragmentation. For small initial densities, no reactions take place and final and initial densities are equal (green dashed line). The first corrections due to reactions are captured by a linked cluster expansion [black dashed line, Eq.~\eqref{eq:linkedcluster}]. The final density is maximal for $\rho=1/2$, beyond which the dense occupation hinders mixing. The limit $\rho=1$ is exactly determined by a recursion argument [gray dotted line, Eq.~\eqref{eq:recursion}].}
    \label{fig:4}
\end{figure}

The limiting case of instant reactions, \mbox{$\lambda/\Gamma \to \infty$}, gives analytical insight into the final fragmented state and we will study it in the remainder of this Letter. Figure~\ref{fig:3} shows  a typical time evolution of an initial state, one event at a time from top to bottom. For instant reactions, there is at most one particle per lattice site and particles no longer move past their neighbors, which restricts transport and leads to faster fragmentation into independent regions. For generality, we consider initial configurations drawn from a Bernoulli distribution with density \mbox{$\rho\in[0,1]$}, which includes the parity projection of the previous Poisson initial condition, for which \mbox{$\rho \leq 1/2$}. Figure~\ref{fig:4} shows the average final density $\rho_\infty$ for different initial densities $\rho$ as obtained from numerical simulation (blue points). For small~$\rho$, the randomly placed particles are unlikely to be close enough to react, so the initial and final densities are approximately equal (green dashed line). As the number of initial particles increases, more reactions take place and the difference between the initial and final density increases. Intriguingly, the final density is not a monotonic function of the initial density but has both a maximum at half filling,~\mbox{$\rho_{\rm max}=1/2$}, and a local minimum at~\mbox{$\rho_{\rm min}\approx0.941$}. This surprising phenomenon indicates that annihilation reactions are very efficient to split the particle configurations into independent clusters already at an early stage of the time evolution, with fewer subsequent possible reactions compared to a homogeneous state with intermediate initial particle density. We note that the nonmonotonic structure is not present for coagulation reactions \mbox{$A + A \rightarrow A$}, which are less efficient to fragment the state space.

To further illustrate the importance of fragmentation in the evolution of the system, we include in Fig.~\ref{fig:4} the final density for a small system of size \mbox{$L=13$} (orange line), obtained by an exact computation of the evolution matrix. This result agrees with the simulation of the extended system up to three decimal places. In particular, for \mbox{$\rho=1/2$} we find \mbox{$\rho_\infty = 0.19290\dots$}, in agreement with the result for an extended system size. This indicates that small values of $L$ are good approximations of infinite lattices, which can thus be considered as ensembles of smaller lattices separated due to fragmentation. The average final density in Fig.~\ref{fig:4} is obtained by evaluating
\begin{equation}
    \rho_{\infty} = \frac{1}{L} \sum_{i, j = 1}^{2^L} n_i U_{i\leftarrow j} \, \rho^{n_j}(1-\rho)^{L-n_j} ,
    \label{eq:poly}
\end{equation}
where $i$ and $j$ label the~$2^L$ possible states that have at most one particle per site, \mbox{$U_{i\leftarrow j}$} is the total time-evolution matrix from a state $j$ at time zero to a state $i$ at infinite time, and~\mbox{$\rho^{n_j}(1-\rho)^{L-n_j}$} is the probability of an initial state $j$ with $n_j$ particles. The time-evolution operator $U$ is expressed in terms of the single-step transition matrix~\mbox{$T_{i\leftarrow j}$} that gives the probability of transitioning from state $j$ to state $i$. We define \mbox{$U = \lim_{k\to\infty} T^{2k}$}, where the factor $2$ ensures the limit is well defined in the presence of eigenvalues $-1$ that would otherwise cause the powers of $T$ to oscillate indefinitely. Such eigenvalues are associated with states containing the oscillatory sequence \mbox{\config{1001}$\leftrightarrows$\config{0110}}, which does not change the particle number. From Eq.~\eqref{eq:poly}, the average final density \mbox{$\rho_\infty$} is a polynomial of order~$L$ in the initial density~$\rho$,
\begin{equation}
    \rho_\infty = \sum_{m=0}^L c^{(L)}_m \rho^m .
\end{equation}
To determine the coefficients $c^{(L)}_m$, we consider the approximated values obtained using \mbox{$U\to T^{2k}$} with finite~$k$ in Eq.~\eqref{eq:poly}, which at large~$k$ differs from the exact value by an exponentially small $k$-dependent transient. Assuming the coefficients $c^{(L)}_m$ are rational numbers, their continued fraction expansion will then consist of the $k$-independent expansion of the exact coefficient followed by a $k$-dependent part. In our numerical calculations, the onset of the $k$-dependent part is marked by an anomalously large expansion term that grows exponentially with $k$, which allows us to obtain the exact value for~$c^{(L)}_m$ without explicitly computing the limit~\mbox{$k\to\infty$}. Numerically, the main hurdle for obtaining the coefficients $c^{(L)}_m$ is the exponential growth of the number of states with system size. We mitigate this by making use of translation and mirror symmetries, which asymptotically reduce the number of states by a factor of $2L$~\cite{oeis24}. 

For small and large initial densities, we obtain analytical results for the final density of infinite systems. Considering first the small-density limit, Eq.~\eqref{eq:poly} implies that only initial configurations with particle number~\mbox{$n_j\leq m$} contribute to the coefficient $c^{(L)}_m$ and that the expansion of the final density~$\rho_\infty$ in powers of~$\rho$ takes the form of a linked cluster expansion. Moreover, as a consequence of the constrained particle mobility, the coefficients~$c_m^{(L)}$ do not change above a critical system size~\mbox{$L_m=3m$} where the lattice becomes effectively infinite, i.e., the evolution of any $m$-particle arrangement is unchanged if placed in a larger lattice. The value $3m$ corresponds to the maximum distance $m$ particles can be separated while still interacting with all neighbors. An explicit calculation as described above for \mbox{$L=21$} gives the first seven coefficients
\begin{equation}
\begin{split}
    \rho_\infty = & \rho -14 \rho^3 + 51 \rho^4 -\frac{214679}{4035}\rho ^5 - \left(214+\frac{1}{\scriptstyle 7 + \frac{1}{2 + \dots}}\right)\rho^6  \\
    &  +\left(1046+\frac{1}{\scriptstyle 2 + \frac{1}{4 + ...}}\right)\rho^7 + O(\rho^8), \label{eq:linkedcluster}
\end{split}
\end{equation}
where we abbreviate the fractions for $c_6$ and $c_7$ as they contain over $100$ terms. The linked cluster expansion up to this order is shown in Fig.~\ref{fig:4} as a dashed black line.

In the opposite limit of a completely filled initial state,  \mbox{$\rho=1$}, an exact value for $\rho_\infty$ (Fig.~\ref{fig:4}, gray dotted line) follows from the observation that empty intervals of size four or longer separate regions that do not interact, as none of the previously discussed ten subsequences that lead to a reaction contains four adjacent empty sites. This leads to a recursive relation for the average final density of the completely filled initial state. Indeed, consider an isolated interval of~$L$ consecutively filled sites and denote the average particle number in its final state by $n_L$. By inspection, we find \mbox{$n_0=0$}, \mbox{$n_1=1$}, \mbox{$n_2=2$}, and \mbox{$n_3=1$}. A possible evolution of a longer interval with \mbox{$L=14$} is illustrated on the right part of Fig.~\ref{fig:3}, where longer intervals fragment into smaller ones by annihilation reactions. An interval of length $L$ contains \mbox{$L-3$} configurations with four consecutively occupied sites \mbox{(\config{1111})} that can annihilate by either an inward or outward hopping step [Figs.~\ref{fig:1}(a) and 1(b)], resulting in two smaller intervals of lengths \mbox{$L_1$} and \mbox{$L_2$} with \mbox{$L_1+L_2=L-4$}. Such an update is shown in the second line of Fig.~\ref{fig:3}, where a filled interval of length \mbox{$L=14$} fragments in two shorter intervals of size \mbox{$L_1=7$} and \mbox{$L_2=3$}. In addition, there are two endpoint configurations \mbox{(\config{0111},} \mbox{\config{1110})} that can annihilate by an outward hop, reducing $L$ by three units and leaving one disconnected particle that does not participate in  further annihilations. Such an update is shown in the third line of Fig.~\ref{fig:3}, where a correlated hopping step takes place at the right boundary of an interval of length \mbox{$L=7$}, leaving a filled interval of length $L=4$ and a single disconnected particle. Thus, an interval of length $L$ has \mbox{$2(L-3)+2 = 2 (L-2)$} equally probable annihilation events, which gives the following recursion relation for the average particle number,
\begin{equation}
    n_L = \frac{1}{L-2} \biggl[\sum_{\scriptscriptstyle L_1+L_2 = L-4} \bigl(n_{L_1} + n_{L_2} \bigr) + n_{L-3} + 1 \biggr]. \label{eq:recursion}
\end{equation}
A completely filled state in a periodic lattice of size~$L$ always annihilates into a stretch of length \mbox{$L-4$}, so that \mbox{$ \rho_\infty = n_{L-4}/L$}. Numerically solving the recursion~\eqref{eq:recursion}, we obtain \mbox{$\lim_{L\to\infty} n_{L-4} / L \approx 0.1807$}, consistent with the numerical result \mbox{$\rho_\infty =  0.1806(1)$} in Fig.~\ref{fig:4}. We have confirmed that Eq.~\eqref{eq:recursion} agrees with the exact value from Eq.~\eqref{eq:poly} for all $L\leq13$.

In conclusion, we have introduced and discussed a lattice model with constrained dynamics and annihilation reactions that displays universal behavior in its approach to the final state. This final state is not empty, and in the limit of instantaneous reactions we obtain exact results for the final density of both finite and infinite lattices. The fragmentation of states into independent regions implies that larger lattices can be considered as ensembles of smaller ones, which allows us to connect finite- and infinite-size lattices with a cluster expansion and obtain exact recursion relations for the completely filled initial state. 
We remark that the model discussed in this work constitutes a minimal model for fragmentation in the dissipative dynamics of a many-body system, and the same effect is present for simple modifications of the dynamics---for example, for variable jump lengths, multiple species, or coalescence reactions---albeit with quantitative changes.  
Possible experimental platforms to realize our proposed dynamics are atoms in optical lattices and excitons in layered van der Waals structures. For the former, both anomalous constrained diffusion~\cite{guardadosanchez20} and annihilation processes in the form of inelastic particle losses~\cite{syassen08} have been realized. For the latter, the anomalous subdiffusion of interlayer excitons is caused by strong repulsive dipole interactions~\cite{tagarelli23}, and two-body annihilation reactions correspond to Auger processes~\cite{erkensten21}.

\begin{acknowledgments}
We thank Grigory Sarnitsky for discussions. This work is supported by Vetenskapsr\aa det (Grant No. 2020-04239). The computations were enabled by resources provided by the National Academic Infrastructure for Supercomputing in Sweden (NAISS) at Linköping University partially funded by the Swedish Research Council through Grant Agreement No. 2022-06725.
\end{acknowledgments}

\bibliography{bib_correlated}

\end{document}